\documentclass[10pt,twocolumn,showpacs,longbibliography,amsmath,amssymb,osajnl,floatfix,superscriptaddress]{revtex4-1}

\usepackage{graphicx}
\usepackage{dcolumn}
\usepackage{bm}
\usepackage{color}
\usepackage{txfonts}
\usepackage{microtype}
\usepackage{epstopdf}
\usepackage{verbatim}
\usepackage[normalem]{ulem}

\begin{document}

%\preprint{AIP/123-QED}

\title{Interferences effects in polarized nonlinear Breit-Wheeler process}

\author{Jing-Jing Jiang}
\affiliation{Department of Physics, Shanghai Normal University, Shanghai 200234, China}
\author{Ya-Nan Dai}
\affiliation{Department of Physics, Shanghai Normal University, Shanghai 200234, China}
\author{Kai-Hong Zhuang}
\affiliation{Department of Physics, Shanghai Normal University, Shanghai 200234, China}
%\author{Karen Z. Hatsagortsyan}
%\author{Christoph H. Keitel}
%\affiliation{Max-Planck-Institut f\"{u}r Kernphysik, Saupfercheckweg 1,
%69117 Heidelberg, Germany}
\author{Yunquan Gao}
\affiliation{College of Physics and Optoelectronic Engineering, Ocean University of China, Qingdao, Shandong 266100, China}
\author{Suo Tang}
\email{tangsuo@ouc.edu.cn}
\affiliation{College of Physics and Optoelectronic Engineering, Ocean University of China, Qingdao, Shandong 266100, China}
\author{Yue-Yue Chen}
\email{yue-yue.chen@shnu.edu.cn}
\affiliation{Department of Physics, Shanghai Normal University, Shanghai 200234, China}
\date {\today}

\begin{abstract}
The creation of polarized electron-positron pairs by the nonlinear Breit-Wheeler process in short laser pulses is investigated using the Baier-Katkov semiclassical method beyond local-constant-field approximation (LCFA), which allows for identifying the interferences effects in the positron polarization.  When the laser intensity is in the intermediate %multiphoton
regime, the interferences of pair production in different formation lengths induce an enhancement of pair production probability for spin-down positrons, which significantly affects the polarization of created positrons.
The polarization features are distinct from that obtained with LCFA, revealing the invalidity of LCFA in this regime. Meanwhile, the angular distribution for different spin states varies, resulting in an angular-dependent polarization of positrons. The average polarization of positrons at beam center is highly sensitive to the laser's carrier-envelope phase (CEP),  which  provides a potential alternative way of determining the CEP of strong lasers. The verification of the observed interference phenomenon is possible for the upcoming experiments.% at FACET II at SLAC and LUXE at DESY.

\end{abstract}

\maketitle
\section{Introduction}
In the presence of a strong electromagnetic field, high-energy photons can  be converted into an electron-positron pair via the nonlinear Breit-Wheeler process \cite{ritus1985quantum,katkov1998electromagnetic,burke1997positron,bamber1999studies,bulanov2013electromagnetic,di2016nonlinear,wan2020high,blackburn2018nonlinear,blackburn2022higher,tang2022fully,meuren2016semiclassical}. This an appealing physical process as it allows for the conversion of light  into particles of matter, and has practical implications for astrophysics, high-energy physics, and laser-plasma interactions \cite{fedotov2023advances,gonoskov2022charged}. The E-144 experiment conducted at SLAC in the 1960s provided the first experimental evidence for the nonlinear Breit-Wheeler  process (NBW) \cite{burke1997positron,bamber1999studies}. Pair production involving multiphoton is observed by colliding a 46.6GeV electron beam with a terawatt laser with peak intensity of $\sim 10^{18}$W/cm$^2$.
The polarized (linear) Breit-Wheeler process has been measured  at the Relativistic Heavy Ion Collider, which  reveals a large fourth-order angular
modulation of the observed yields \cite{adam2021measurement}.
The advancement of high-power laser technology \cite{danson2015petawatt,vulcan,ELI} has sparked interest in revisiting the E-144 experiment with lasers of higher intensity.
The upcoming experiments at FACET II at SLAC \cite{salgado2021single} and LUXE at DESY \cite{abramowicz2021conceptual} aim to explore NBW in the tunneling regime, where the production rate follows an exponential scaling law of $\exp(-\pi E_{\text{cr}}/E)$ \cite{schwinger1951gauge}. Here  $E$ is the electric field strength and $E_{\text{cr}}=m^2/|e|$ the critical field of quantum electrodynamics (QED), with $e<0$ and $m$ being the electron charge and mass, respectively. However, the experiments will be initially conducted in the intermediate laser intensity regime, where the transition from multiphoton to tunneling regime is expected to occur. The separation of regimes is determined by the nonlinear parameter $a_0=|e|E/m\omega_{0}$, with $\omega_{0}$ being the laser carrier frequency.

In the tunneling regime of $a_0\gg1$, the formation length of pair creation is much smaller than the spatial and temporal inhomogeneities of the laser field \cite{katkov1998electromagnetic,ritus1985quantum,wiedemann2003synchrotron}. The probability can be calculated under the locally constant field approximation (LCFA) \cite{ritus1985quantum}.
Recently, various LCFA probabilities have been derived  incorporating electron-positron spins and photon polarization \cite{Seipt2018,seipt2020spin,chen2022electron,
li2019ultrarelativistic,li2020polarized,torgrimsson2021loops},
and has been applied to QED-simulations codes for studying polarization effects in strong-field QED processes \cite{song2022dense,zhuang2023laser,li2018ultrarelativistic,
li2020polarized,chen2019polarized,Wan_2019,dai2022photon,
li2020production,li2022helicity,chen2022electron,seipt2019ultrafast}. Ultrarelativistic positrons (electrons) beam with high degree of polarization can be generated via NBW in asymmetric laser fields \cite{chen2019polarized}, as a result of spin preferences parallel (antiparallel) to the instantaneous quantization axis during creation. However, the current or upcoming experiments are  performed with intermediate values of laser intensity ($a_0\sim1$),
where the formation length of pair production $l_f\sim \lambda/a_0$ is comparable with the laser field inhomogeneities. In this case, interferences effects arising from the pair creation process may significantly modify the positrons polarization.%\cite{olugh2020asymmetric,page1959electron,moortgat2008polarized}.

The interference effects has been investigated in several works \cite{wistisen2014interference,meuren2016semiclassical,titov2013breit,titov2016quantum, ilderton2020coherent,akkermans2012ramsey,jansen2017strong,granz2019electron,ILDERTON2020135410}. It has been reported that the interferences occurring at macroscoptically separated space-times points could induce interference structures in the momentum distribution of the the final particles \cite{meuren2016semiclassical}, and interference between  pair creation channels  of different photon number combinations could enhance the yield of pairs in bichromatic fields \cite{jansen2016strong}. Meanwhile, the carrier-envelope phase (CEP) effects are pronounced for a few-cycle pulses with moderate laser intensity \cite{meuren2015polarization}, which induces shifts of spectra and angular distribution due to interferences \cite{meuren2016semiclassical,jansen2016strong,titov2013breit,titov2016quantum}.
Moreover, the polarization effects in pair production has been investigated in the transition regime ($a_0\sim1$) \cite{jansen2016strong,krajewska2012breit,tang2021pulse}. The oscillation pattern in angular spectra of positrons \cite{krajewska2012breit} and positron yield \cite{wistisen2020numerical_nbw,TangPRD056003} can be enhanced by tuning photon polarization.  The relevance of spin effects on NBW has been investigated by comparing the production of spinor and scalar particles \cite{villalba2013photo,jansen2017strong}. However, a comprehensive study of the positron polarization in this regime has yet to be conducted.

To improve the accuracy of simulation codes in the intermediate intensity regime, a method based on locally monochromatic approximation (LMA) has been developed taking into account interference effects at the scale of the laser wavelength \cite{bamber1999studies,heinzl2020locally,hartin2018strong}.
In contrast to LCFA,  LMA is more suitable for exploring the transition region %\blue{\sout{, especially}} 
for planewave-like fields with slowly varying envelope. 
While for a short laser pulse, LMA would underestimate the pair production as it loses the high-frequency components induced by the finite-pulse envelope. In fact, these overlooked components have the potential to effectively enhance pair production~\cite{tang2022fully}.
While for a short laser pulse, the fully spin-resolved pair-creation probability could be obtained using the semiclassical operator method of Baier \textit{et al.} \cite{katkov1998electromagnetic,baier1998electromagnetic}, which is applicable to non-plane-wave background. The probability associated with particular spin states and photon polarization can be calculated by coherently integrating over the electron trajectory. The numerical results obtained for plane waves exhibit good agreement with the predictions of the Volkov state approach \cite{volkov1935class}, implying that the semiclassical method is capable of describing the polarized Breit-Wheeler process in the intermediate intensity regime \cite{wistisen2020numerical_ncs,wistisen2020numerical_nbw,nielsen2022high}.

In this paper, we investigate the NBW in a linearly polarized ultrastrong laser pulse using the semiclassical method beyond LCFA. It allowed us to
identify interference effects in positrons polarization and angular distribution. We show that the positrons polarization exhibits unique features due to interferences that cannot be captured by LCFA. The spectra and angular distribution of final particles are depend sensitively on the spin states, leading to an angular dependent polarization of positrons. By cutting the positrons distributed at small angle region, one could obtain polarized positrons with polarization degree $\sim 25\%$.  Moreover, the polarization degree is proportional to CEP, which is attractive for relativistic positron generation with controlled polarization and CEP measurement in strong laser fields.
\begin{figure*}
    \includegraphics[width=0.9\textwidth]{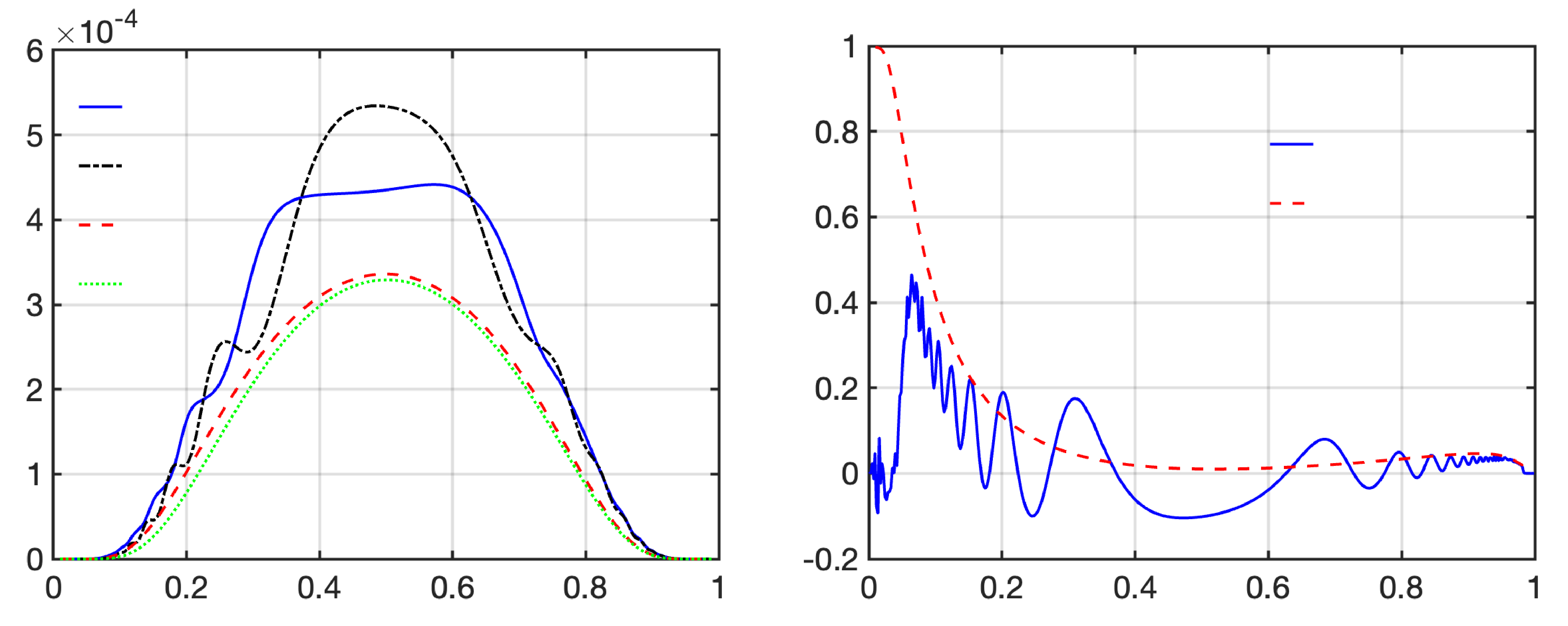}
    \begin{picture}(300,20)
    \put(110,175){\fontsize{11pt}{\baselineskip}\selectfont (a)}
    \put(189,175){\fontsize{11pt}{\baselineskip}\selectfont (b)}
    \put(32,12){\fontsize{11pt}{\baselineskip}\selectfont$\delta_+$}
    \put(268,12){\fontsize{11pt}{\baselineskip}\selectfont$\delta_+$}
    \put(-86,100){\rotatebox{90}{\fontsize{11pt}{\baselineskip}\selectfont$dP/d\delta_+$}}
    \put(140,112){\rotatebox{90}{\fontsize{11pt}{\baselineskip}\selectfont$\zeta_f^+$}}
    \put(-41,170){Semi. $\uparrow$}
    \put(-41,153){Semi. $\downarrow$}
    \put(-41,136){LCFA $\uparrow$}
    \put(-41,118){LCFA $\downarrow$}
    \put(307,159){Semi.}
    \put(307,143){LCFA}
    \end{picture}
    \caption{(a) The pair production probability  $dP/d\delta_+$ versus $\delta_+$ after averaging over photon polarization: for spin-up positrons using semiclasscial approach (red-dotted) or the LCFA approach  (blue-solid); for spin-down positrons using semiclasscial approach (green-dot-dashed) or the LCFA approach (black-dashed). (b) The spectra of positron polarization with semiclasscial approach (blue-solid) or the LCFA  approach  (red-dashed). The arrow denotes the positron spin with respect to $\hat{\mathbf{y}}=(0,1,0)$.}
    \label{Fig.1d}
\end{figure*}
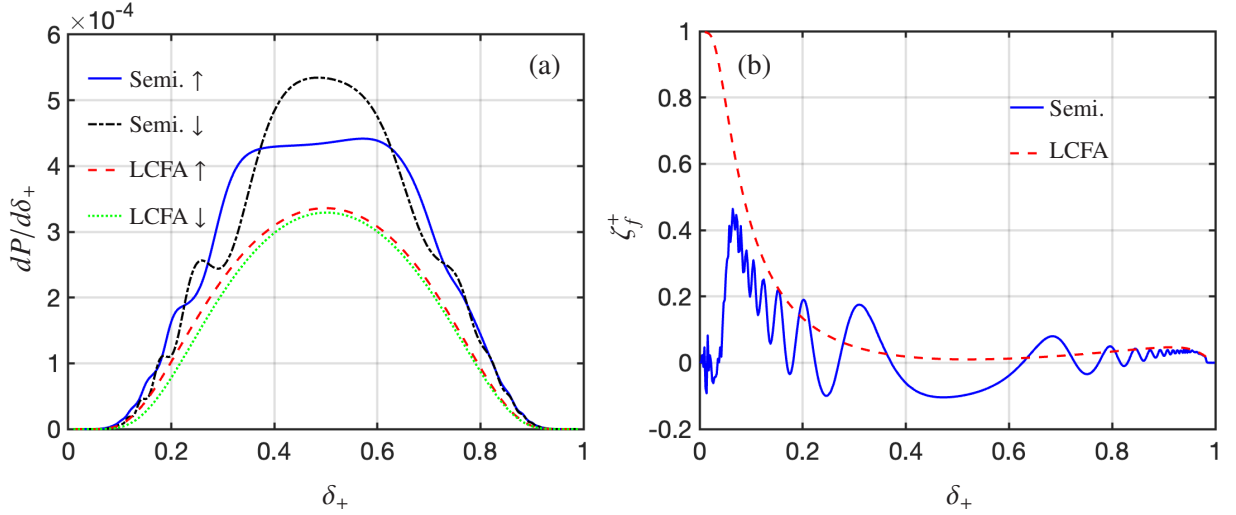

\section{Theoretical framework}\label{Theoretical framework}
%%%%%%%%%%%%%%
\subsection{Semiclassical approach beyond LCFA}
We consider the head-on collision of a 93.5GeV photon with a linearly polarized laser pulse propagating along $-z$ direction. XCELS \cite{XCELS} envisions advanced accelerator complexes capable of achieving particle energies in the 1-10 TeV range, which could be utilized to generate gamma quanta with energies of hundreds of GeV through Compton backscattering \cite{jansen2016strong,wistisen2020numerical_nbw}. The background field takes the form of $E_x=E_{0}\sin^{4}\left(\varphi/2N\right)\cos\left(\varphi+\varphi_{0}\right)$ if $\varphi\in\{0,2\pi N\}$ and $E_x=0$ otherwise, where $\varphi$ is the laser phase of its carrier frequency $\omega_{0}$.
The peak intensity is $I\sim 10^{18}\text{W/cm}^2 (a_0=1)$, wavelength of the laser  $\lambda_0=800$nm, $N=5$ corresponding to a pulse duration of $\sim3T_0$ with period $T_0$, and CEP $\varphi_0=0$. As the emitted photon number for a single electron is much smaller than 1 ($N_{\gamma}\sim\alpha a_{0}\tau_{p}/T_{0}\ll1$  \cite{ritus1985quantum,di2012extremely}), the photon emissions of produced particles can be disregarded. Otherwise, the produced pairs could undergoes further radiation in the field, which could change the spin states due to radiative polarization and radiative corrections. In this case, the polarization features predicted by pair production probability itself may not be accurate.

For the semiclassical approach without LCFA, the angle-resolved spectral probability for electron can be obtained by integrating the trajectories of electrons  with different final momentum $\bm{p}_-^f=(p^f_-\cos\psi\sin\theta,p^f_-\sin\psi\sin\theta,p^f_-\cos\theta)$. One should go through all possible choices of angles and electron momenta, obtain the trajectories for each final momentum by solving Lorentz equation with initial electron momentum of $\bm{p}_-^i=\bm{p}_-^f$. Note that, in the cases where the initial and final momenta are different due to radiation reaction or laser configuration, one need to resort the obtained probabilities with the final momentum $\bm{p}_-^f$. The Lorentz equation is solved using the Runge-Kutta algorithm with a time step $dt=10^{-19}$s, which is small enough to ensure the convergence of integrals. The integrals are evaluated with summations over finite differenced points $N=t_f/dt$.  At each discrete point,
the time-dependent momenta and coordinates of the electrons are substituted into the angle-resolved spectrum for electrons \cite{belkacem1985theory,wistisen2020numerical_nbw}. The  rapidly oscillating function $e^{i(\varepsilon_{-}/\varepsilon_{+})k_{\mu}x_{-}^{\mu}}$ in the integrals is split into real and imaginary parts, which are calculated separately in each time step.
Since the energy (momentum) of ultrarelativistic electron and positron pairs is correlated, the angle-resolved spectral probability for positrons can be obtained through the electron trajectory:
\begin{align}\label{CD}\nonumber
\frac{d^2P}{d\varepsilon_{+}d\Omega} & =\frac{e^{2}}{\left(2\pi\right)^{2}}\frac{\varepsilon_{-}^{2}\mathbb{N}^{2}}{\omega\omega^{\prime2}}\left|\phi_{-}^{\dagger}\left\{ \varepsilon_{-}\omega\left[\mathbf{n}J-\mathbf{I}\right]\cdot\left(\boldsymbol{\epsilon}\times\mathbf{n}\right)-\right.\right.\\
 & \left.\left.i\boldsymbol{\sigma}\cdot\left[\boldsymbol{\epsilon}\omega mJ-2\varepsilon_{-}^{2}\mathbf{K}-\varepsilon_{-}\omega\mathbf{n}\left(\boldsymbol{\epsilon}\cdot\mathbf{I}\right)\right]\right\} \phi_{+}\right|^{2},
\end{align}
where
\begin{align*}
\mathbb{N} & =\frac{1}{\sqrt{4\varepsilon_{-}\varepsilon_{+}\left(\varepsilon_{-}+m\right)\left(\varepsilon_{+}+m\right)}},
\end{align*}
\begin{align*}
\mathbf{I}&=\int\frac{\mathbf{n}\times\left[\left(\mathbf{n}-\mathbf{v}_{-}\right)\times\mathbf{a}_{+}\right]}{\left(1-\mathbf{n}\cdot\mathbf{v}_{-}\right)^{2}}e^{i\frac{\varepsilon_{-}}{\varepsilon_{+}}k_{\mu}x_{-}^{\mu}}dt,
\end{align*}
\begin{align*}
J&=\int\frac{\mathbf{n}\cdot\mathbf{a}_{-}}{\left(1-\mathbf{n}\cdot\mathbf{v}_{-}\right)^{2}}e^{i\frac{\varepsilon_{-}}{\varepsilon_{+}}k_{\mu}x_{-}^{\mu}}dt,
\end{align*}
\begin{align*}
\mathbf{K}&=\int\left(\frac{[\mathbf{a}_{-}\left(\boldsymbol{\epsilon}\cdot\mathbf{v}_{-}\right)+\mathbf{v}_{-}\left(\boldsymbol{\epsilon}\cdot\mathbf{a}_{-}\right)]}{\left(1-\mathbf{n}\cdot\mathbf{v}_{-}\right)}+\right.\\&\left.+\frac{[\mathbf{v}_{-}\left(\boldsymbol{\epsilon}\cdot\mathbf{v}_{-}\right)\left(\mathbf{n}\cdot\mathbf{a}_{-}\right)]}{\left(1-\mathbf{n}\cdot\mathbf{v}_{-}\right)^{2}}\right)e^{i\frac{\varepsilon_{-}}{\varepsilon_{+}}k_{\mu}x_{-}^{\mu}}dt.
\end{align*}
Here $k_{\mu}=\omega m\left\{ 1,\mathbf{n}\right\}$ is the four-momentum of the parent photon with $\mathbf{n}=(0,0,1)$ being the direction of motion of the parent photon,  $\omega^{\prime}=\frac{\varepsilon_{-}}{\varepsilon_{+}}\omega$ with $\varepsilon_{+}$ ($\varepsilon_{-}$) being the energy of the created positron (electron) and $\omega=(\varepsilon_{+}+\varepsilon_{-})/m$ the energy of the parent photon.  The photon polarization vector is defined as $\bm{\epsilon}=\{\mathbf{e}_\parallel,\mathbf{e}_\perp\}$ with
\begin{align}
\mathbf{e}_\parallel&=\hat{\mathbf{y}}\times\mathbf{n}, \mathbf{e}_\perp=\mathbf{n}\times \mathbf{e}_\parallel,
\end{align}
where $\mathbf{\hat{y}}=(0,1,0)$ being the unit vector along magnetic field direction. For the considered setup, $\mathbf{e}_\parallel$ and $\mathbf{e}_\perp$ are the laser electric and magnetic field direction, respectively.
$x_{+}^{\mu}=\left\{ t,\mathbf{r}\left(t\right)\right\}$, $\mathbf{v}_{-} $ and $\mathbf{a}_{-} $ are the four-coordinate, velocity and acceleration of the produced electron, respectively.
$\phi_{-} $ and $ \phi_{+} $ are the bispinor of the electrons and positrons, respectively, and $\boldsymbol{\sigma}$ are the Pauli matrices.  If the $y$-axis is selected as the spin quantization axis, then $\phi_{-} =\left(\frac{1}{2}-i\frac{1}{2}, \frac{1}{2}+i\frac{1}{2}\right)^{T} $ corresponds to the electron's spin-up state, and $ \phi_{+} =\left(-\frac{1}{2}+i\frac{1}{2},\frac{1}{2}+i\frac{1}{2}\right)^{T} $ represents the positron's spin-up state. %Note that the angle-resolved spectrum for positron can be obtained straightforwardly from Eq. (\ref{CD}) by using $\varepsilon_+=\omega-\varepsilon_-$

\subsection{Spin-resolved LCFA approach}
To illustrate the interference effects, we also investigated NBW using the spin-resolved LCFA probability. The LCFA assumed the formation length for pair production is much smaller than the wavelength and period of the field, which allows one to regard the field inside the formation length as constant, i.e. ignore the interference within a formation length,  and drop out the rapidly oscillating terms related to interferences between different formation lengths. Therefore, a comparison between semiclassical and LCFA results could reveal the role of interference. The spin-resolved LCFA probability can be written as $P_{\pm}=(P \pm \Delta P)/2$, where $P$ is the total unpolarized probability~\cite{10.1063/5.0159963}:
\begin{align}
\frac{d^{2}P}{d \varepsilon_{+} d\Omega}&=\frac{\alpha}{2\pi } \delta_{+}^{2}\int\frac{ d\varphi}{\omega_{0}}
\left[ \nu\left(\frac{\varepsilon_{-}}{\varepsilon_{+}}+\frac{\varepsilon_{+}}{\varepsilon_{-}} \right) + \mu\right]\textrm{Ai}(\nu)\,,
\end{align}
and $\Delta P$ is the difference in probability due to creating the positron into $\pm
\mathbf{\hat{y}}$ spin-polarization
state:
\begin{align}
\frac{d^{2}\Delta P}{d \varepsilon_{+} d\Omega}=\frac{\alpha}{2\pi } \delta_{+} \int\frac{d\varphi}{\omega_{0}}  \sqrt{\mu}\textrm{Ai}'(\nu) \textrm{sgn}(E_{x}) \,,
\end{align}
where %$\varepsilon_{+}$ ($\varepsilon_{-}$) is the energy of the created positron (electron), $\omega=\varepsilon_{-}+\varepsilon_{+}$, and
 $\delta_{+}=\varepsilon_{+}/\omega$ is the fraction of the energy taken by the positron from the parent photon, and
 \begin{equation}
 \mu=[\chi_{\ell} \delta_{+}(1-\delta_{+})]^{-2/3}\,,~\nu=\mu(1+ \varepsilon^{2}_{+}\sin^{2}\theta / m^{2})\,,
 \end{equation}
with $\chi_{\ell}=\omega E/E_{cr}$.

%fully spin-resolved Monte Carlo method. This method employs LCFA probabilities derived with semiclassical approach of Baier, Katkov and Strakhovenko \cite{zhuang2023laser,li2018ultrarelativistic,li2020polarized,chen2019polarized,Wan_2019,dai2022photon,li2020production,li2022helicity}.
%At each simulation step the pair production is determined by the total LCFA probability, and the photon energy by the spectral probability, using the common algorithms. The spin of produced pairs is determined by the spin-resolved probabilities \cite{chen2022electron} according to the stochastic algorithm and instantaneously collapsed into one of its basis states defined with respect to the instantaneous spin quantization axis (SQA).
%After the pair production, the electrons (positrons) motion in the external field is governed by the Lorentz force, while spin precession is described by Bargmann-Michel-Telegdi (BMT) equation \cite{Bargmann_1959,Walser_2002}. To compare with the semiclaasical approach without LCFA, we consider a monochromatic gamma-ray beam consists of $\sim10^6$ photons, which has a cylindrical form with radius $w_e=\lambda_0$, length $L_e=5\lambda_0$ and initial energy of $93.5$GeV. The beam density has a transversely Gaussian and longitudinally uniform distribution.
%Assessing the differences between the semiclassical approaches with and without LCFA allows us to evaluate the impact of interferences.

\section{numerical calculations}

%The effects of interference on polarization is illustrated in Fig. \ref{Fig.1d}.
Using the semiclassical method and the parameters given in Sec. \ref{Theoretical framework} A, one could obtain the spectrum for different particles' polarization states (see Appendix B).
For a more intuitive perspective,
 we average over photon polarizations and sum over electron polarizations to specifically examine the spin-resolved probability of positron with respect to $\mathbf{\hat{y}}$, see  Fig.~\ref{Fig.1d} (a).
It is evident that the probability for positron spin-down is higher than that for positron spin-up as a result of asymmetry of the short laser pulse, while the LCFA underestimates this difference in the probabilities for positron spin-up and spin-down.
In the semiclassical method, interference peaks appear in the spectra [Fig.\ref{Fig.1d} (a)].  The interferences lead to an enhancement of pair production probabilities and a more pronounced asymmetry between different polarization states of positrons.
Counterintuitively, the probability for positron spin-down exceeds spin-up.
This is because interference effects arising from pair production  at negative half-cycles with $B_y<0$ are more significant than that at positive half-cycles, leading to a notable enhancement of pair production probability with negative positron spin, \emph{i.e.}, $\zeta^+_{f}<0$.

%the probabilities for different approaches do not coincide [see Fig. \ref{Fig.1d} (b)].  The probabilities for positron spin-up and spin-down are roughly the same for LCFA results, and the former is slightly larger than the later as a result of asymmetry of the short laser pulse. While in the case without LCFA, interference peaks appear in the spectra [Figs.\ref{Fig.1d} (a) and (b)].  The interferences lead to an enhancement of pair production probabilities and a more pronounced asymmetry between different polarization states of positrons. Counterintuitively, the probability for positron spin-down exceeds spin-up with respect to $\mathbf{\hat{y}}$. This is because interference effects arising from pair production  at negative half-cycles with $B_y<0$ are more significant than that at positive half-cycles, leading to a notable enhancement of pair production probability with negative positron spin ($\zeta^+<0$).

The disparities in probability give rise to distinct polarization characteristics across different approaches.
For the LCFA approach, the polarization of positron decreases with the increase of positron energy, which coincides with that in a constant field with $B_y>0$. The average polarization of the positron beam is $3.8\%$.
However, the polarization undergoes significant changes in the results from the semiclassical method, in which interferences are introduced [see Fig. \ref{Fig.1d} (b)]. Surprisingly, the average polarization becomes $-1.5\%$. The positron polarization $\zeta_f^+$ decreases with large oscillations at small position energy $\delta_+$, exhibits a negative dip around $\delta_+=0.5$ with $|\zeta_f^+|$ reaching to $\sim10\%$, and finally diminishes to zero when $\delta_+>0.6$. The LCFA approach fails to capture the interference structures, resulting in an overestimation of polarization for low energy positrons. Conversely, it underestimates the polarization near the density peak. Therefore, for $a_0=1$, the LCFA performs  poorly in accurately describing the polarization of the created particle.

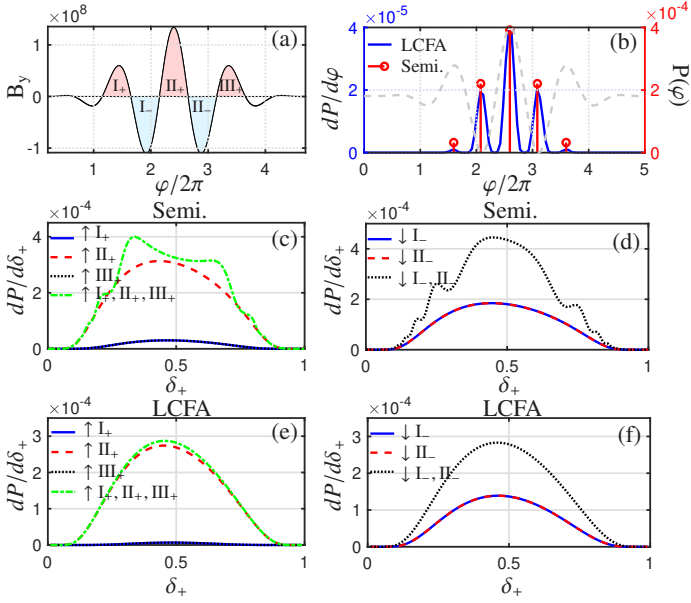
\begin{figure}[t]
    \includegraphics[width=0.5\textwidth]{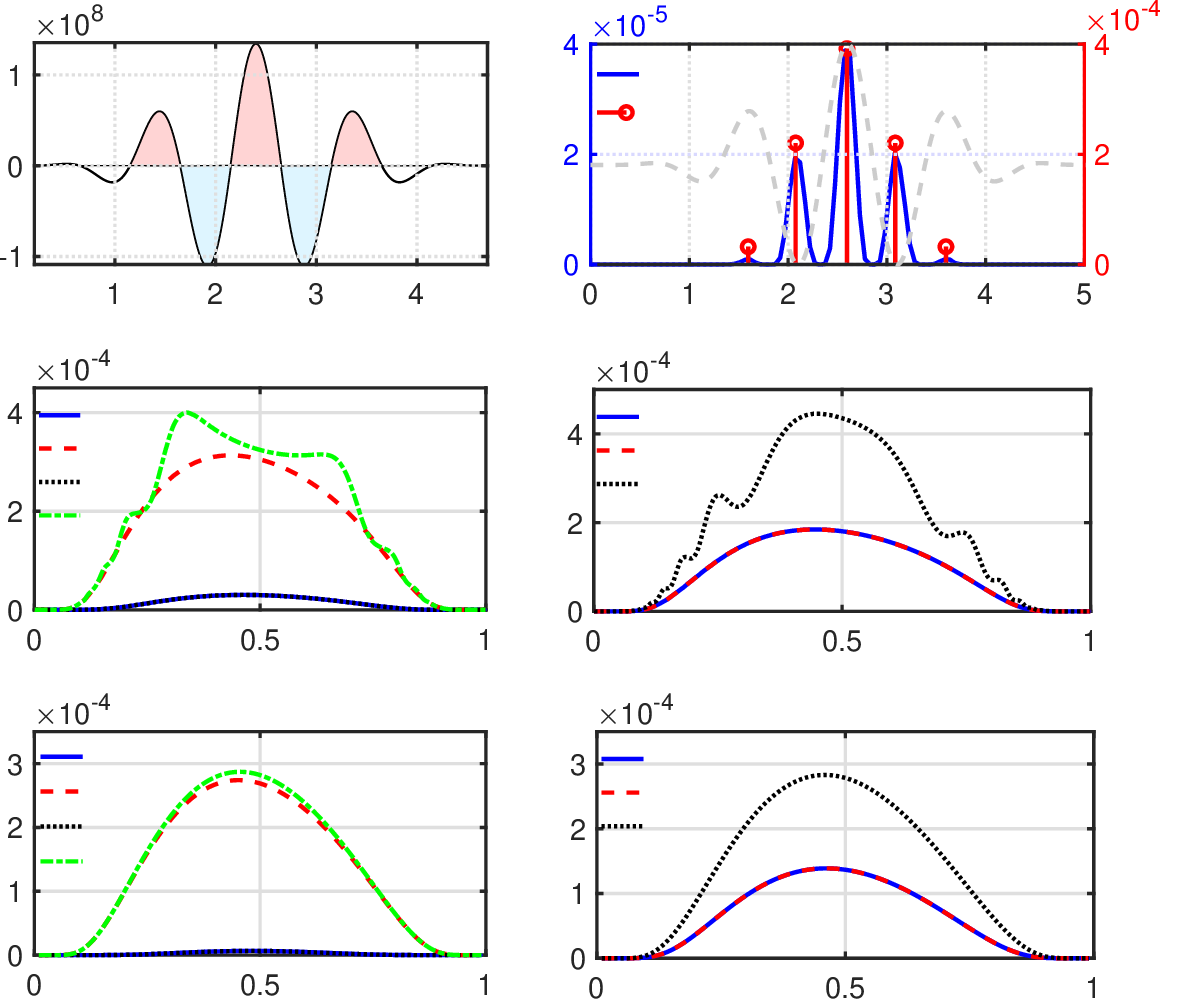}
 \begin{picture}(300,20)
     \put(47,155){Semi.}
     \put(172,155){{Semi.}}
     \put(47,80){{LCFA}}
     \put(172,80){{LCFA}}
    \put(92,220){(a)}
 \put(220,219){(b)}
 \put(92,144){(c)}
 \put(221,143){(d)}
 \put(92,71){(e)}
 \put(223,70){(f)}
 \put(48,165){$\varphi/2\pi$}
    \put(-8,200){\rotatebox{90}{$\mathrm{B_y}$}}    
    \put(32,203){\fontsize{6.5pt}{\baselineskip}\selectfont$\mathrm{I}_+$}
    \put(52,203){\fontsize{6.5pt}{\baselineskip}\selectfont$\mathrm{II}_+$}
    \put(72,203){\fontsize{6.5pt}{\baselineskip}\selectfont$\mathrm{III}_+$}
    \put(42,195){\fontsize{6.5pt}{\baselineskip}\selectfont$\mathrm{I}_-$}
    \put(62,195){\fontsize{6.5pt}{\baselineskip}\selectfont$\mathrm{II}_-$}
    \put(172,165){$\varphi/2\pi$}
    \put(111,188){\rotatebox{90}{$dP/d\varphi$}}
     \put(141,218){\fontsize{7.0pt}{\baselineskip}\selectfont LCFA}
     \put(141,210){\fontsize{7.0pt}{\baselineskip}\selectfont Semi.}
    \put(242,213){\rotatebox{270}{P$(\varphi)$}}
    \put(52,89){$\delta_+$}
    \put(-8,118){\rotatebox{90}{$dP/d\delta_+$}}
    \put(20,147){\fontsize{7.0pt}{\baselineskip}\selectfont$\uparrow\mathrm{I}_+$}
    \put(20,139){\fontsize{7.0pt}{\baselineskip}\selectfont$\uparrow\mathrm{II}_+$}
    \put(20,131){\fontsize{7.0pt}{\baselineskip}\selectfont$\uparrow\mathrm{III}_+$}
 \put(20,124){\fontsize{7.0pt}{\baselineskip}\selectfont$\uparrow\mathrm{I}_+,\mathrm{II}_+,\mathrm{III}_+$}
 \put(178,89){$\delta_+$}
   \put(112,118){\rotatebox{90}{$dP/d\delta_+$}}
    \put(140,146){\fontsize{7.0pt}{\baselineskip}\selectfont$\downarrow\mathrm{I}_-$}
 \put(140,138){\fontsize{7.0pt}{\baselineskip}\selectfont$\downarrow\mathrm{II}_-$}
 \put(140,130){\fontsize{7.0pt}{\baselineskip}\selectfont$\downarrow\mathrm{I}_-,\mathrm{II}_-$}
     \put(-8,44){\rotatebox{90}{$dP/d\delta_+$}}
    \put(21,73){\fontsize{7.0pt}{\baselineskip}\selectfont$\uparrow\mathrm{I}_+$}
    \put(21,65){\fontsize{7.0pt}{\baselineskip}\selectfont$\uparrow\mathrm{II}_+$}
    \put(21,57){\fontsize{7.0pt}{\baselineskip}\selectfont$\uparrow\mathrm{III}_+$}
 \put(21,49){\fontsize{7.0pt}{\baselineskip}\selectfont$\uparrow\mathrm{I}_+,\mathrm{II}_+,\mathrm{III}_+$}
 \put(52,13){$\delta_+$}
     \put(112,44){\rotatebox{90}{$dP/d\delta_+$}}
    \put(141,72){\fontsize{7.0pt}{\baselineskip}\selectfont$\downarrow\mathrm{I}_-$}
 \put(141,64){\fontsize{7.0pt}{\baselineskip}\selectfont$\downarrow\mathrm{II}_-$}
 \put(141,56){\fontsize{7.0pt}{\baselineskip}\selectfont$\downarrow\mathrm{I}_-,\mathrm{II}_-$}
  \put(179,13){$\delta_+$}
    \end{picture}
    \caption{(a) The magnetic field and labels for different half-cycles.  (b) Probability density d$P$/d$\varphi$ at laser phase $\varphi$ using LCFA (blue-solid) and probability at each half-cycles using semiclassical approach (red-circle). Spectral probabilities for spin-up positrons in the half-cycle I$_+$ (blue-solid), II$_+$ (red-dashed), III$_+$ (black-dotted) and the superposition of I$_+$, II$_+$  and III$_+$ (green-dot-dashed) for: (c) semiclassical approach and (e) LCFA approach. Spectral probabilities for spin-down positrons in the half-cycle I$_-$ (blue-solid), II$_-$ (red-dashed), and the superposition of I$_+$ and II$_+$  (black-dotted) for: (d) semiclassical approach and (f) LCFA approach. Here the spin-up and spin-down is defined with respect to $\mathbf{\hat{y}}=(0,1,0)$. The parameters are same as Fig. 1.}
        \label{Fig.Bypm}
\end{figure}

For the LCFA approach, the positron polarization arises due to the asymmetry of the ultrashort laser pulse and the preference for pair production in spin-up positrons,  %with respect to $\mathbf{\hat{y}}$, %\sout{the instantaneous SQA. Since},
 as the LCFA probability is solely depend on the local quantum parameter $\chi_\gamma\propto a_0\omega$ and the positrons are mostly produced at the intense positive half-cycle II$_+$. Since the pair production is more probable for spin-up positron, the positrons produced at the half-cycle II$_+$ are more likely to be polarized with $\zeta_f^+>0$ [Fig. \ref{Fig.Bypm} (b)]. However, due to the cancellation of polarization effect between positive and negative half-cycles, the average polarization of positron in an ultrashort laser pulse is rather small ($\sim 3.8\%$).

When interferences are taken into account, the density of positrons with $\zeta_f^+<0$ gets larger than that of $\zeta_f^+>0$, indicating more pairs are produced at the negative half-cycles I$_-$ and II$_-$ than the positive peak II$_+$ as shown in Fig. \ref{Fig.Bypm}.
This is because interferences play an essential role in determining the spin-resolved pair production probabilities. The influences of interferences within each laser cycle  and between macroscopically separated cycles are illustrated in Fig. \ref{Fig.Bypm} (c) and (d). It can be seen that the positrons with spin parallel with $\mathbf{\hat{y}}$ are mostly comes from pair production at the pulse peak II$_+$. The positive half-cycles I$_+$ and III$_+$ also contribute but with a much lower amplitude [Fig. \ref{Fig.Bypm} (c)]. Due to the interferences within each laser cycle, the pair production amplitude at  I$_+$, II$_+$ and III$_+$ is all  enhanced compared to the results obtained from the LCFA [Fig. \ref{Fig.Bypm} (e)]. Meanwhile, at $a_0\lesssim1$, the coherence length is of the order of the period of the electron trajectory, which means that the process is formed along all multicycle dynamics, i.e., effectively the formation length is the whole trajectory length. In this case, the interferences between cycles also make an important role.
When considering pair productions during all the positive half-cycles, interference structures appear in the spectrum.

\begin{figure}[b]
    \includegraphics[width=0.45\textwidth]{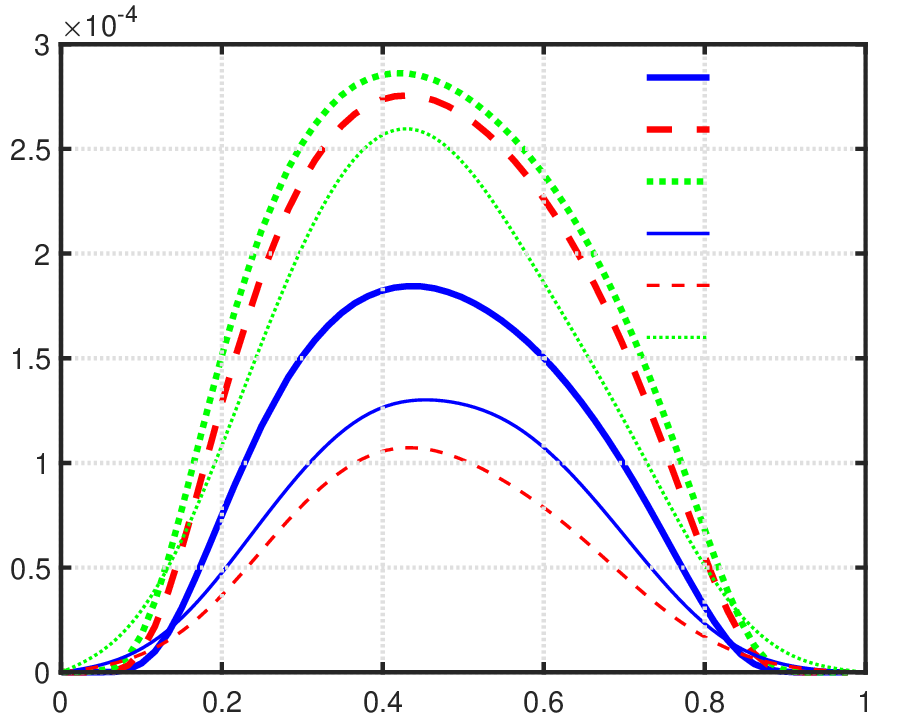}
    \begin{picture}(300,20)
    \put(187,181){\fontsize{7pt}{\baselineskip}\selectfont Semi.$\downarrow$ 0}
    \put(187,169){\fontsize{7pt}{\baselineskip}\selectfont Semi.$\downarrow$ $\pi$/2}
    \put(187,156){\fontsize{7pt}{\baselineskip}\selectfont Semi.$\downarrow$ $\pi$/10}
    \put(187,143){\fontsize{7pt}{\baselineskip}\selectfont LCFA $\downarrow$ }
    \put(187,130){\fontsize{7pt}{\baselineskip}\selectfont LCFA $\downarrow$ $\pi$/2}
    \put(187,117){\fontsize{7pt}{\baselineskip}\selectfont LCFA $\downarrow$ $\pi$/10}
    \put(120,12){\fontsize{9pt}{\baselineskip}\selectfont$\delta_+$}
    \put(0,103){\rotatebox{90}{\fontsize{9pt}{\baselineskip}\selectfont$dP/d\delta_+$}}
    \end{picture}
    \caption{Spectral probabilities for spin-down positrons in the half-cycle II$_-$ for laser pulse with  CEP $\varphi_0=0$ (blue solid line), $\varphi_0=\pi/2$ (red danshed line), and $\varphi_0=7\pi/10$ (green dotted line): for semiclassical approach (thick line) and LCFA approach (thin line). The rest parameters are same with Fig. \ref{Fig.1d}.}
  \label{Fig.cep2}
\end{figure}
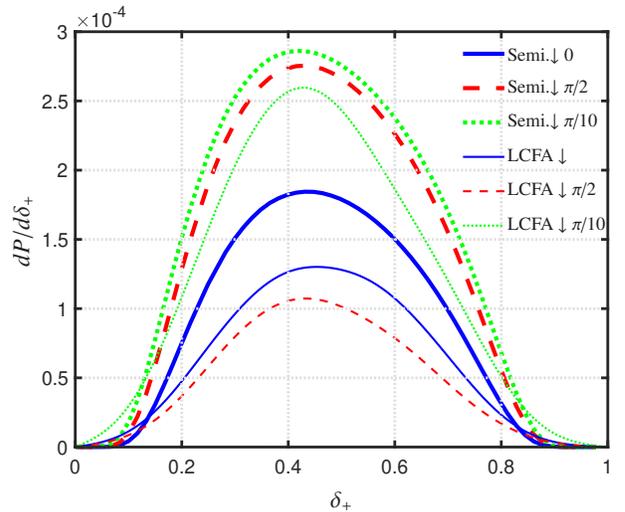

\begin{figure}[b]
    \includegraphics[width=0.5\textwidth]{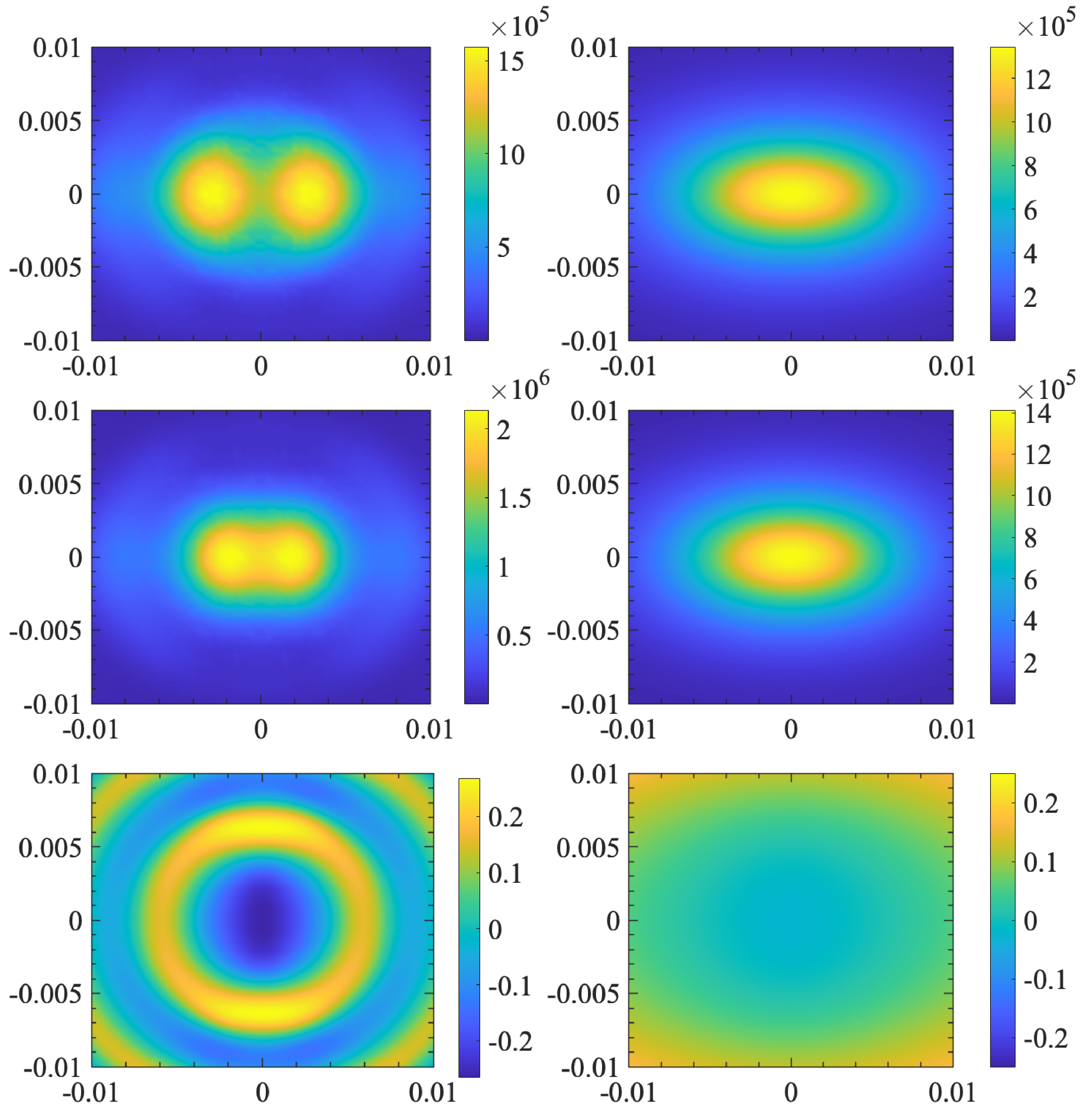}
    \begin{picture}(300,15)
    \put(52,270){{Semi.}}
    \put(172,270){{LCFA}}
    \put(25,255){\color{white}(a)}
    \put(25,170){\color{white}(c)}
    \put(25,87){\color{white}(e)}
    \put(58,10){$\theta_x$}
    \put(183,10){$\theta_x$}
    \put(0,228){\rotatebox{90}{$\theta_y$}}
     \put(0,144){\rotatebox{90}{$\theta_y$}}
      \put(0,60){\rotatebox{90}{$\theta_y$}}
    \put(152,255){\color{white}(b)}
    \put(152,170){\color{white}(d)}
    \put(152,87){\color{white}(f)}
    \end{picture}
    \caption{Angular distribution of positron density $d^2P/d\theta_x/d\theta_y$ (rad$^{-2}$) versus $\theta_x$ (mrad) and $\theta_y$ (mrad): for positron spin-up (a) and spin-down (c) calculated with semiclassical approach without LCFA; for positron spin-up (b) and spin-down (d) calculated with LCFA approach. Angular distribution of positron polarization $\zeta_f^+$ versus $\theta_x$ (mrad) and $\theta_y$ (mrad) calculated with semiclassical approach without LCFA (e) and LCFA probability (f). The parameters are same as Fig. 1.}
       \label{Fig.txty}
\end{figure}
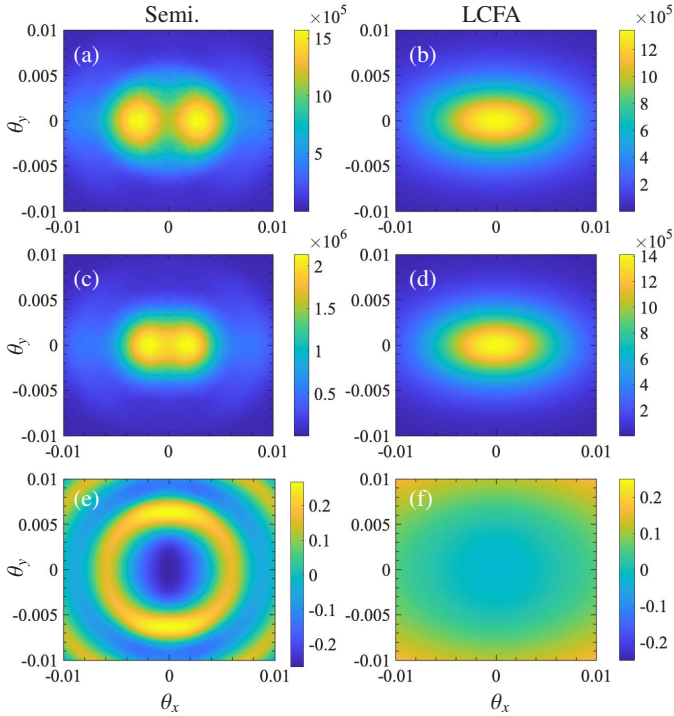

 In contrast, the spin-down positrons mainly come from pair production at negative half-cycles I$_-$ and II$_-$, which contribute equally to positron spectrum. The interferences within each half-cycle I$_-$ and II$_-$ increase the corresponding probability compared with the LCFA results in Fig. \ref{Fig.Bypm} (e). More importantly,
the constructive interferences arising from pair production occurring at I$_-$ and II$_-$ result in an enhancement of positrons density with $\zeta_f^+>0$ and $\delta_+\approx0.5$ [Fig. \ref{Fig.Bypm} (d)].
With respect to $\mathbf{\hat{y}}$, the pair production is more probable for spin-down positron around $\delta_+\in(0.4,0.6)$. However, outside of this region, probability for spin-up dominates over spin-down. This is because the half-cycle II$_+$ is more intense than the negative half-cycles [Fig. \ref{Fig.Bypm} (a)],
%pair production at positive peak II$_+$ involves higher order harmonics than negative half-cycles.
there is a higher number of photons $n$ involved in pair production during $\mathrm{II}_+$
compared to the other cycles.
The width of the spectrum increases as the harmonic order $n$ increases \cite{ivanov2005complete}. The maximum of  $n$ are determined by the field intensity where pairs are produced. The interference between different half-cycles only affects the amplitude of harmonics through an interference pattern, but it doesn't change the harmonic order.
Consequently, the spectrum for spin-down positrons exhibits a broader width compared to the negative half-cycles, leading to a positive polarization outside the range of $\delta_+\in(0.4,0.6)$. It is important to note that the observed interference effects, including interference structures in spectrum and negative polarization, predominantly arise from pair production events that occur at macroscopically separated formation regions. The interferences within each cycle, on the other hand, only result in an overall enhancement of the respective probability.

The interferences within each half-cycle always result in an increase in pair production probability compared with LCFA, see Fig. \ref{Fig.cep2}.
This is because the laser field in a finite time domain has a wide spectrum. The high frequency components make a significant contribution to the pair production probability, which are ignored by LCFA.
In contrast, the interferences between macroscopically separated cycles could be destructive or constructive determined by  the phase differences. See the discussion in Ref. \cite{ilderton2019exact}. In our case, the interferences patten of two adjacent negative (positive) half-cycles can be quantitatively interpreted by the interferences from two trajectory segments that have the same velocity and acceleration. In this case, the integral over the entire trajectory can be estimated with $P\propto |e^{i(\varepsilon_{-}/\varepsilon_{+})k_{\mu}x_{1-}^{\mu}}+e^{i(\varepsilon_{-}/\varepsilon_{+})k_{\mu}x_{2-}^{\mu}}|^2$, where $x_{1-}^{\mu}$ and $x_{2-}^{\mu}$ are four-coordinates for the two trajectory segments with a $2\pi$ difference. The destructive interferences occur when
$n\pi=(\varepsilon_{-}/\varepsilon_{+})\omega \Delta t(1-\beta_-\cos\theta)\approx m^{2}\pi/\delta_{+}\left(1-\delta_{+}\right)\omega\omega_{0}$, resulting in the emergence of valleys in the spectrum at $\delta_{+}=n\omega\omega_{0}\pm\sqrt{-4m^{2}n\omega\omega_{0}+n^{2}\omega^{2}\omega_{0}^{2}}/2n\omega\omega_{0}$ \cite{jackson1999classical,katkov1998electromagnetic,di2012extremely}. Similarly, constructive peaks emerge when $2n\pi=(\varepsilon_{-}/\varepsilon_{+})\omega \Delta t(1-\beta_-\cos\theta)$. Therefore, the probability of pair production is enhanced mainly due to the interference within each half-cycles, and interference between cycles results in destructive valleys and constructive peaks superimposed on the enhanced probability.\\

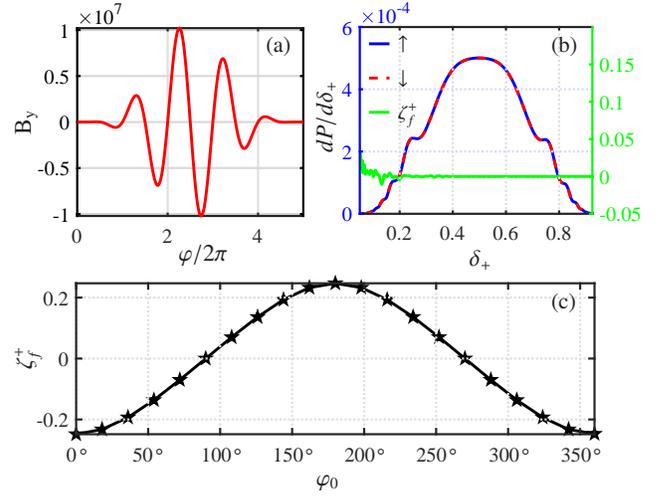
\begin{figure}[t]
    \includegraphics[width=0.46\textwidth]{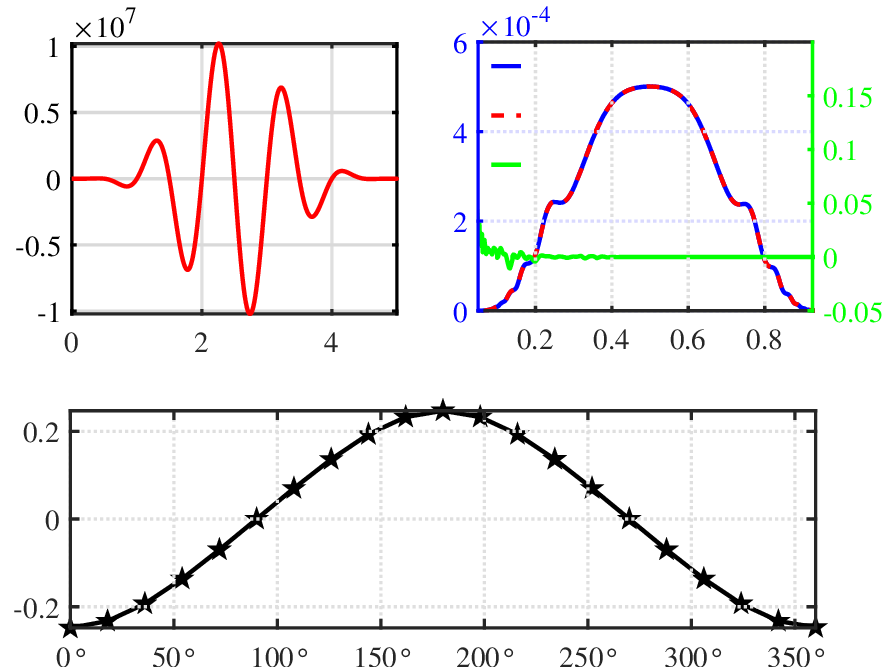}
    \begin{picture}(300,20)
    \put(96,176){\fontsize{8.5pt}{\baselineskip}\selectfont (a)}
    \put(204,176){\fontsize{8.5pt}{\baselineskip}\selectfont (b)}
    \put(204,79){\fontsize{8.5pt}{\baselineskip}\selectfont (c)}
    \put(1,145){\rotatebox{90}{\fontsize{9.0pt}{\baselineskip}\selectfont$\mathrm{B_y}$}}
    \put(63,97){\fontsize{9.0pt}{\baselineskip}\selectfont$\varphi/2\pi$}
    \put(173,95){\fontsize{9.0pt}{\baselineskip}\selectfont$\delta_+$}
    \put(115,138){\rotatebox{90}{\fontsize{9.0pt}{\baselineskip}\selectfont$dP/d\delta_+$}}
    \put(1,56){\rotatebox{90}{\fontsize{9.0pt}{\baselineskip}\selectfont$\zeta_f^+$}}
    \put(146,177){\fontsize{7.6pt}{\baselineskip}\selectfont$\uparrow$}
    \put(146,164){\fontsize{7.6pt}{\baselineskip}\selectfont$\downarrow$}
    \put(146,151){\fontsize{7.6pt}{\baselineskip}\selectfont $\zeta_f^+$}
    \put(115,12){\fontsize{9.0pt}{\baselineskip}\selectfont$\varphi_0$}
    %\put(232,154){\rotatebox{90}{\fontsize{10.0pt}{\baselineskip}\selectfont$\overline{\zeta}_f^+$}}
    \end{picture}
     \caption{ The variation of the magnetic field with respect to the laser phase $\varphi$ (a) and the spectral probabilities for spin-up and spin-down positrons, with respect to $\mathbf{\hat{y}}$ (b)  for a CEP of $\varphi_0=\pi/2$. (c) The scaling law of average polarization $\overline{\zeta}_f^+$ within $\theta\in [-0.02\text{mrad}, 0.02\text{mrad}]$ with CEP $\varphi_0$. The rest parameters are same with Fig. \ref{Fig.1d}.}
        \label{Fig.cep}
\end{figure}

The angular distributions of the positrons in different spin states are illustrated in Fig. \ref{Fig.txty}. For the semiclassical approach without LCFA, the distribution of positrons exhibits two symmetric circular regions around $\theta_x=0$. The separation is induced by longitudinal acceleration appeared in $J\propto \mathbf{\dot{v}}_{z}\sim v_xB_y$. At $\theta_x\approx0$, the positron density  decreases as $J$ approaches zero, and increases to a maximum at $\theta_x\approx \pm a_0/2\gamma\sim \pm3\times10^{-6}$rad. The separation between the two regions is larger for the positrons with spin-up than spin-down with respect to $\mathbf{\hat{y}}$ [Fig. \ref{Fig.txty} (a) and (c)]. This because the final momentum of positrons is determined by the laser intensity at creation, i.e. $p_f^x\approx eA(\varphi_+)$ with $\varphi_+$ being the creation phase. The  positrons with $\zeta^+_{f}>0$ are mostly generated in the high laser intensity region, consequently, they have higher transverse momentum $p_x$ compared to the positrons with $\zeta^+_{f}<0$.
Therefore, the small angle region is dominated by the spin-down positrons created at negative half-cycles, while the large angle region is determined by the spin-up positrons created at positive half-cycles, leading to an angle dependent polarization [see Fig. \ref{Fig.txty} (e)]. By collecting the positrons at small angle regions, one could obtain polarized positrons with average polarization reaching to $\sim26\%$.
In contrast, the angular distribution obtained with LCFA approach exhibits an elliptical shape for different spin states [Fig. \ref{Fig.txty} (b) and (d)]. It is because positrons are most created at laser peaks with $p_x(\varphi_+)\approx0$ [Fig.\ref{Fig.Bypm} (b)], and consequently have a small $\theta_x$ out of laser. Moreover, the interference ring structure, which is present in the semiclassical approach, disappears in the LCFA approach. The probability for spin-up always dominates over spin-down, which is more pronounced for low-energy positrons deflected to large angle region [Fig. \ref{Fig.txty} (f)].

For the polarized Breit-wheeler process in short laser pulse, it is possible to identify the CEP effects in the positron polarization. In the case of $\varphi_0=\pi/2$  [Fig. \ref{Fig.cep} (a)], the pair production probabilities for different spin states are equivalent, leading to a vanished polarization throughout the spectrum [Fig. \ref{Fig.cep} (b)].
This is because the ultrashort laser pulse becomes symmetric when $\varphi_0=\pi/2$. Polarization effects in different half-cycles cancel out due to the symmetry of the laser field. By collecting the small angle positrons, we obtain the scaling law of average polarization on the CEP of laser [Fig. \ref{Fig.cep} (c)]. The polarization increases monotonically from -25\% to 25\%  when the CEP increases from 0 to $\pi$, and decreases from 25\% to -25\% when the CEP increases from $\pi$ to $2\pi$. The sensitive dependence of polarization on CEP provides an potential way of detecting strong laser's CEP.

\section{Conclusion}
In this paper, we have investigated the polarization effects of the NBW using a short laser pulse in the intermediate intensity regime. By comparing the semiclassical approach beyond LCFA and the LCFA approach, we scrutinized the validity of the LCFA in predicting the polarization of created pairs. The positrons produced in a short pulse exhibit a net polarization as a result of the asymmetry of the background field.  However, the polarization features obtained from different approaches can be completely different, even when considering the same asymmetric field.
The LCFA approach demonstrates a positive polarization due to the dominant influence of the peak positive half-cycle, and the polarization degree decreases monotonically with the increase of positron energy. %However, the interference effects are ignored in the simulations employing LCFA, which are essential for an accurate description of polarization effects when the laser field strength is on the order of $a_0\sim O(1)$.
When taking interference effects into account, it is observed that the density of positrons with $\zeta^+<0$ surpasses that of positrons with $\zeta^+>0$ around $\delta_+=0.5$, leading to a negative polarization in the spectrum center. The variation of polarization feature is result from the interferences arising from separated formation lengths. The pair production occurring at negative half-cycles interferences constructively, leading to an enhancement of pair production for positrons with $\zeta^+<0$. %The pair production occurring at the positive peak contributes to a wide spectrum due to the multiple photon absorption.
Interferences structures appear in spectrum for spin-up positrons but without a substantial enhancement of probability. Consequently, a negative polarization emerges around $\delta_+=0.5$, which is contrast to the predictions of LCFA. Meanwhile, positrons presents an angular dependent polarization. The positrons deflected towards the small angle region exhibit a negative polarization of $-25\%$ for $\varphi_0=0$. This negative polarization increases as the carrier-envelope phase is increased before reaching $\varphi_0=\pi$. However, after $\varphi_0=\pi$, the trend is reversed and the polarization starts to decrease. The sensitive dependence of polarization on CEP provides an alternative way of generating and manipulating the positron polarization, as well as CEP measurement in strong laser fields.

\section*{Appendix A: Comparison between semiclassical and QED approaches}
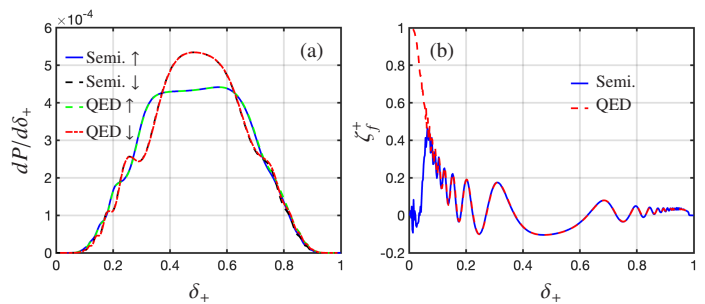
\begin{figure}[b]
    \includegraphics[width=0.5\textwidth]{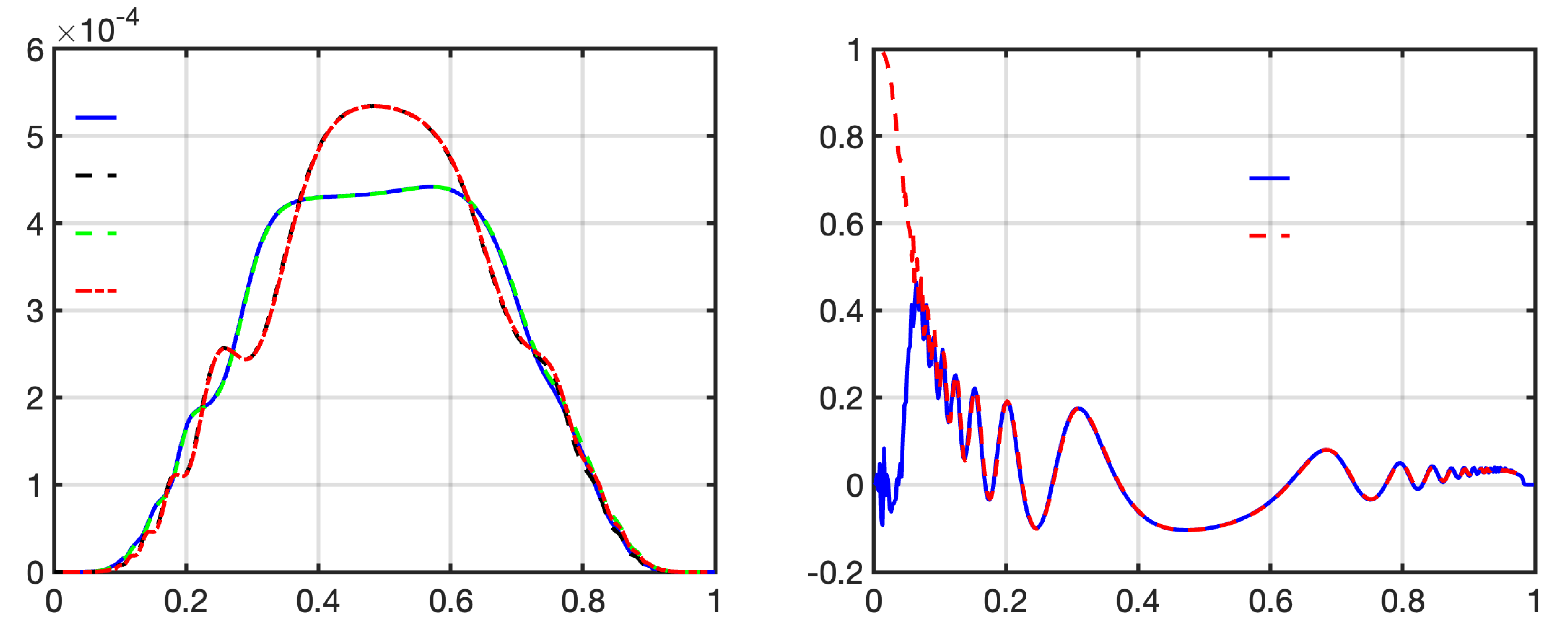}
\begin{picture}(300,20)
    \put(101,105){\fontsize{8pt}{\baselineskip}\selectfont (a)}
    \put(150,105){\fontsize{8pt}{\baselineskip}\selectfont (b)}
    \put(59,14){\fontsize{8.5pt}{\baselineskip}\selectfont$\delta_+$}
    \put(192,14){\fontsize{8.5pt}{\baselineskip}\selectfont$\delta_+$}
    \put(-8,61){\rotatebox{90}{\fontsize{8.5pt}{\baselineskip}\selectfont$dP/d\delta_+$}}
    \put(20,103){\fontsize{6.5pt}{\baselineskip}\selectfont Semi. $\uparrow$}
    \put(20,94){\fontsize{6.5pt}{\baselineskip}\selectfont Semi. $\downarrow$}
    \put(20,85){\fontsize{6.5pt}{\baselineskip}\selectfont QED $\uparrow$}
    \put(20,75){\fontsize{6.5pt}{\baselineskip}\selectfont QED $\downarrow$}
    \put(121,73){\rotatebox{90}{\fontsize{8.5pt}{\baselineskip}\selectfont$\zeta_f^+$}}
    \put(213,94){\fontsize{6.5pt}{\baselineskip}\selectfont Semi.}
    \put(213,85){\fontsize{6.5pt}{\baselineskip}\selectfont QED}
    \end{picture}
        \caption{ (a) The pair production probability  $dP/d\delta_+$ versus $\delta_+$: for spin-up positrons using semiclasscial approach (blue-solid) or the QED approach  (green-dashed); for spin-down positrons using semiclasscial approach (black-dashed) or the QED approach (red-dot-dashed). (b) The spectra of positron polarization with semiclasscial approach (blue-solid) or the QED  approach  (red-dashed). }
        \label{Fig.QED_semi}
\end{figure}

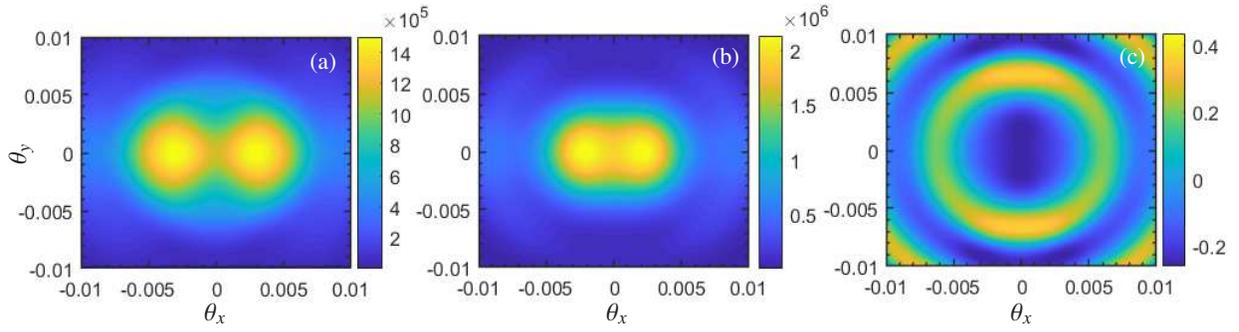
\begin{figure*}[]
    \includegraphics[width=0.9\textwidth]{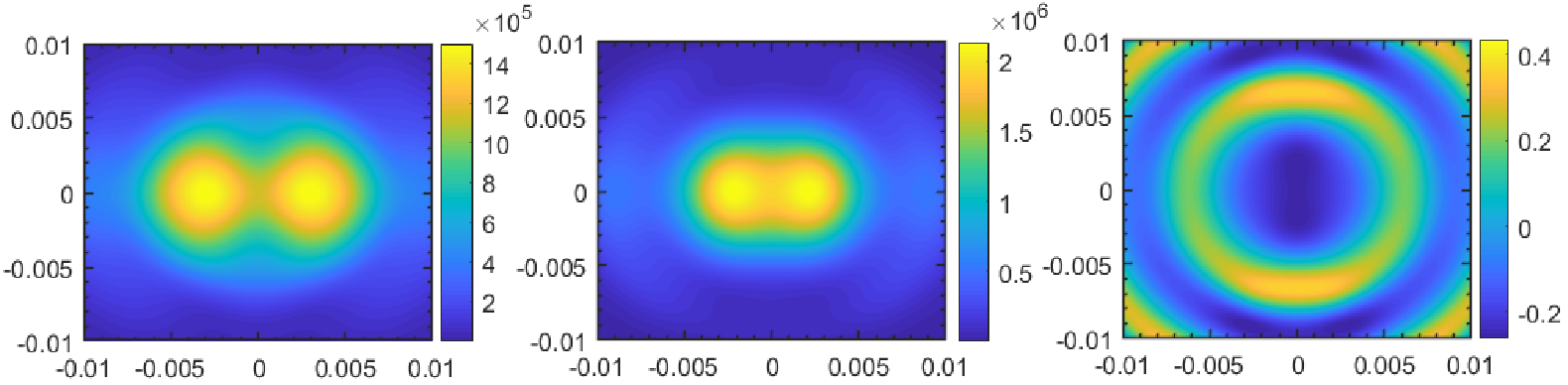}
       \put(-348,88){\color{white}(a)}
       \put(-196,90){\color{white}(b)}
       \put(-42,90){\color{white}(c)}
       \put(-388,-7){\fontsize{10pt}{\baselineskip}\selectfont$\theta_x$}
       \put(-236,-7){\fontsize{10pt}{\baselineskip}\selectfont$\theta_x$}
       \put(-83,-7){\fontsize{10pt}{\baselineskip}\selectfont$\theta_x$}
       \put(-462,52){\rotatebox{90}{\fontsize{10pt}{\baselineskip}\selectfont$\theta_y$}}
     \caption{Angular distribution of positron density $d^2N/d\theta_x/d\theta_y$ (rad$^{-2}$) versus $\theta_x$ (mrad) and $\theta_y$ (mrad) calculated with  QED approach: for positron spin-up (a) and spin-down (b). (c) Angular distribution of positron polarization $\zeta_f^+$ versus $\theta_x$ (mrad) and $\theta_y$ (mrad) calculated with QED approach. }
        \label{Fig.QED_semi2}
\end{figure*}

In this section, we compared the semiclassical results with the full QED calculation derived from the standard S-matrix approach, see Eqs~(20) and (21) in Ref.~\cite{tang2022fully}.
The spin-resolved probabilities from the QED calculation display excellent agreement with the semiclassical approach, markedly differing from the results yielded by the LCFA [Fig. \ref{Fig.QED_semi} (a)]. The polarizations obtained from different approaches also coincide with each other except for the low energy region of $\delta_+<0.6$  [Fig. \ref{Fig.QED_semi} (b)]. The relative difference between the QED and Semi. results within the range of $\delta_+ \in [0.4,0.6]$ is represented by $R_{i}=\left|\left(\frac{dP_{\text{QED}}^{i}}{d\delta_{+}}-\frac{dP_{\textrm{Semi.}}^{i}}{d\delta_{+}}\right)/\frac{dP_{\text{QED}}^{i}}{d\delta_{+}}\right|< 1\%$ with $i\in\left\{ \uparrow,\downarrow\right\}$. Due to the excellent agreement within this range, the average polarization remains consistent at approximately the level of -1.5\% over the entire spectrum. %consistent across different approaches. % The discrepancy at the low-energy end of the spectrum is a result of the failure of energy-momentum conservation at small $\delta_+$, making it invalid to calculate the positron spectrum using the electron trajectory. Moreover,
 The QED calculation is derived using the parameter $s=kp_+/kk'$, while semiclassical method is derived using the parameter $\delta_+=\varepsilon_+/\omega$.
In the case of large positron energy $\varepsilon_+$, these two parameters are equivalent, which allows for a comparison of results obtained using different approaches.
However, in the low-energy regime, the approximation of $s=kp_+/kk'\approx \delta_+$ may not hold, which could be  responsible for the discrepancy between different approaches. Moreover, the discrepancy could be induced by the numerical error of semiclassical method. To accurately describe the spectrum in the low-energy region, it is necessary to extend the trajectory integral to infinity. However, this requires an infinite number of discrete points for numerical integration, which is not possible in practice. Therefore, a finite integration range could result in discrepancies in the results.
 %The discrepancy could be induced by the fact that the QED calculations based on the fraction  $s=kp_+/kk'$, and the approximation $s\approx\delta_+$ do not strictly hold for low-energy positrons.
This deviation can be considered inconsequential when evaluating the average polarization due to the negligible positron density in this region.
Furthermore, the angular distribution of positrons obtained using the semiclassical approach closely matches that of the QED approach  [Figs. \ref{Fig.QED_semi2} (a) and (b)], as well as the interferences rings exhibited in the polarization distribution [Fig. \ref{Fig.QED_semi2} (c)].

\section*{Appendix B: photon polarization}
\begin{figure}
    \includegraphics[width=0.45\textwidth]{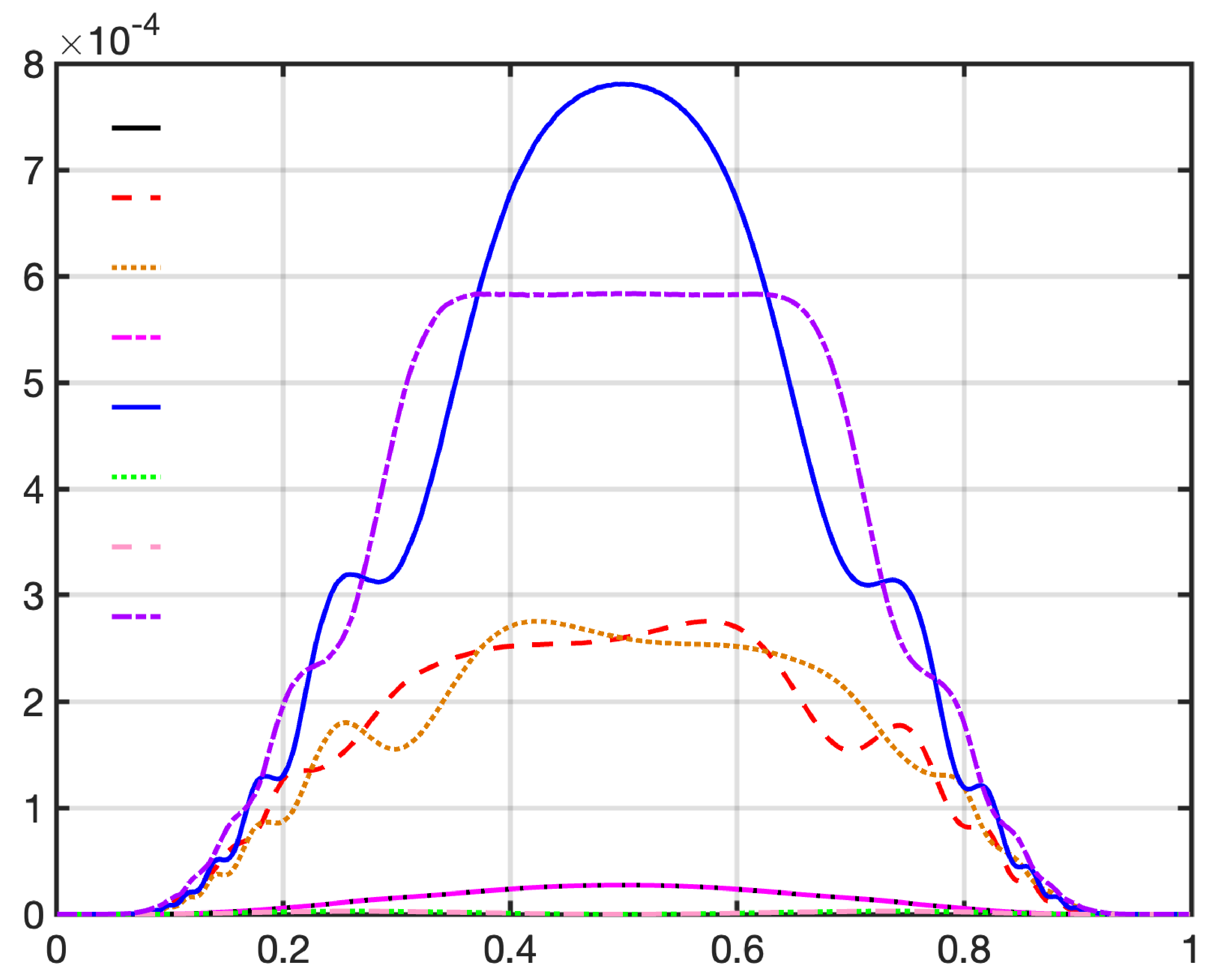}
    \begin{picture}(300,20)
    \put(45,178){$\uparrow\downarrow \parallel$}
    \put(45,164){$\uparrow\uparrow \parallel$}
    \put(45,151){$\downarrow\downarrow \parallel$}
    \put(45,138){$\downarrow\uparrow \parallel$}
    \put(45,127){$\uparrow\downarrow \perp$}
    \put(45,114){$\uparrow\uparrow \perp$}
    \put(45,99){$\downarrow\downarrow \perp$}
    \put(45,86){$\downarrow\uparrow\perp$}
    \put(120,12){\fontsize{9pt}{\baselineskip}\selectfont$\delta_+$}
    \put(0,103){\rotatebox{90}{\fontsize{9pt}{\baselineskip}\selectfont$dP/d\delta_+$}}
    \end{picture}
    \caption{ The  spectral probability $dP/d\delta_+$ versus $\delta_+$ for different electron and positron spins and photon polarizations. The first and second arrow denote the electron and positron spin with respect to $\hat{\mathbf{y}}=(0,1,0)$, respectively, and the last symbol denotes the photon polarization $\epsilon\in\{\mathbf{e}_\parallel,\mathbf{e}_\perp\}$.  }
    \label{Fig.1a}
\end{figure}
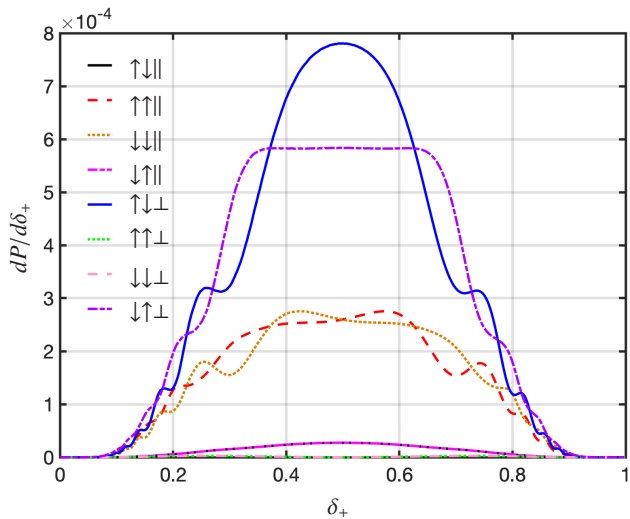

The semiclassical method allows one to obtain the spectrum for arbitrary particles' polarization states, see Fig. \ref{Fig.1a}.  All curves are symmetric about $\delta_{+}$ = 0.5 except $\uparrow\uparrow\parallel$ and $\downarrow\downarrow\parallel$, and the curves for  $\uparrow\uparrow\parallel$ and $\downarrow\downarrow\parallel$ are mirror images of each other about $\delta_{+}$ = 0.5.
The probability is dominated by $dP_{\uparrow\downarrow \mathbf{e}_\perp}$, which corresponds to the scenario where the incoming photon is polarized perpendicularly to the laser polarization along the direction of magnetic field $\mathbf{\hat{y}}=(0,1,0)$, while the final electron has spin-up and the positron has spin-down with respect to  $\mathbf{\hat{y}}$.
In the case that the photon is initially polarized along magnetic field direction, the  spin-resolved probabilities and polarization of created positron are roughly the same with the unpolarized case.  On the other hand, when the photon has initial polarization along the electric field direction, the positron's polarization is negligible, making it unsuitable for studying positron polarization.

%\section*{Appendix C: Enhancement of pair production probability induced by interference}

{\it Acknowledgement:}
We gratefully acknowledge helpful discussions with Karen Z. Hatsagortsyan. This work is supported by the National Natural Science Foundation of China (Grants No. 12074262 and Grant No.12104428), the National Key R\&D Program of China (Grant No. 2021YFA1601700), and the Shanghai Rising-Star Program.
\bibliography{lpp}

\end{document}